\documentclass[10pt,superscriptaddress,twocolumn,amsmath,amssymb,aps,prb]{revtex4}
\usepackage{mathrsfs}
\usepackage{graphicx}
\usepackage{dcolumn}
\usepackage{bm}
\usepackage{array}
\usepackage{booktabs}

\newcommand{\be}{\begin{equation}}
\newcommand{\ee}{\end{equation}}
\newcommand{\bea}{\begin{eqnarray}}
\newcommand{\eea}{\end{eqnarray}}

\begin{document}

\title{$p+ip$-wave pairing symmetry at type-II van Hove singularities}
\author{Yinxiang Li}\email{yinxiangl@hotmail.com}
\affiliation {Tin Ka-Ping College of Science, University of Shanghai for Science and Technology, Shanghai, 200093, China}
\author{Xiaotong Yang}
\affiliation {Tin Ka-Ping College of Science, University of Shanghai for Science and Technology, Shanghai, 200093, China}

\date{\today}

\begin{abstract}
Based on the random phase approximation calculation in two-orbital honeycomb lattice model, we investigate the pairing symmetry of Ni-based transition-metal trichalcogenides by electron doping access to type-II van Hove singularities (vHs). We find that chiral even-parity $d+id$-wave (E$_{g}$) state is suppressed by odd-parity $p+ip$-wave (E$_{u}$) state when electron doping approaches the type-II vHs. The type-II vHs peak in density of states (DOS) enables to strengthen the ferromagnetic fluctuation, which is responsible for triplet pairing. The competition between antiferromagnetic and ferromagnetic fluctuation results in pairing phase transition from singlet to triplet pairing. The Ni-based transition-metal trichalcogenides provide a promising platform to unconventional superconductor emerging from electronic DOS.
\end{abstract}


\maketitle
\section{Introduction}\label{sectioni}
In the past decades, novel topological states of quantum matters are the active and attractive topics in condensed matter physics. The discovery of quantum spin Hall in HgTe has boosted the search of two-dimension (2D) topological insulator\cite{Bernevig,MK}. After that, three-dimension (3D) topological insulators and topological semimetals have also been verified in both theoretical calculation and experiment\cite{HJZhang,YXia,LFu,Hsieh,ZJWang,HMWeng,BQLv}. Meanwhile, topological superconductor\cite{NNHao,XXWu1,XXWu2,DFWang,ZBYan,QYWang,RXZhang,XXWu3} has also attracted tremendous attentions due to the development of topological insulator. The topological superconductor with particle-hole symmetry can host Majorana zero mode which may be potentially employed in realizing topological quantum computation\cite{Kitaev,Freedman}. Especially, it is interesting that 2D chiral $p+ip$ wave topological superconductor in magnetic vortex cores can host Majorana zero modes\cite{Read}. Recently, the outstanding studies are experimental evidences for Majorana bound states in an iron-based superconductor FeTe$_{0.55}$Se$_{0.45}$ by scanning tunneling spectroscopy\cite{DFWang,LYKong,SYZhu}. Thus, searching for intrinsic topological superconductor is the active and prominent field in condensed matter physics.

2D materials include not only quantum spin Hall insulators and Chen insulators, but also unconventional superconductor as doped twisted bilayer graphene\cite{YCao1,YCao2,Yankowitz,XLu}.  Among them, ternary transition-metal phosphorus trichalcogenide (TMPT) compounds APX$_{3}$ (A=3d transition metals; X=chalcogens) have attracted enormous attention due to antiferromagnetic (AF) ordering as a hint for significant electronic correlations\cite{Wildes,Chittari,Sivadas,JULee,CTKuo}. By suppressing AF order with external pressure, superconductivity emerges in iron-based TMPT compounds such as FePSe$_{3}$, with the highest T$_{c}$ found at about 5.5 K\cite{YGWang}. The crystal structure of the TMPT family APX$_{3}$ consists of edge shared AX$_{6}$ octahedral complexes and P2 dimers. Transition metal atoms are arranged in a hexagonal lattice. In the octahedral crystal field, five 3d orbitals of transition-metal atoms split into high-energy e$_{g}$ orbitals and low-energy $t_{2g}$ orbitals. For FePX$_{3}$ with Fe$^{2+}$ ions ($d^{6}$), it is an ideal system to study the high-to-low spin-state transition by pressure. For the case of NiPX$_{3}$ with Ni$^{2+}$ $d^{8}$ filling configuration, $t_{2g}$ bands are fully occupied while $e_{g}$ bands are half filled and dominate the spectral weight near the Fermi level. Theoretical calculations suggest that charge doping can suppress magnetic order, and superconductivity can eventually be achieved for NiPS$_{3}$\cite{YHGu}.

With increasing electron doping, the system far away from half filling and approaches type-II van Hove singularities (vHs)\cite{HYao} along $\Gamma$-M and $\Gamma$-K high symmetry line. For 2D superconductor, a vHs in the density of states (DOS) was proposed to drive a substantial enhancement of interaction effects and to promote the unconventional superconductor\cite{HYao,Nandkishore,WSWang,Kiesel,TXMa,ZYMeng,XChen,LDZhang,CCLiu,Isobe,DDSante,XXWu4}. In general, superconductors with type-I vHs (the saddle points locate at time reversal invariant momentums (TRIMs)) favor singlet pairing. For type-II vHs superconductors (the saddle points at general k points), the triplet pair can compete with singlet pair\cite{HYao}. By random phase approximation (RPA) calculation in honeycomb lattice, we find that the $p+ip$-wave (E$_{u}$) pairing is enhancing and then overcomes an I-wave (A$_{2g}$) state and a chiral d-wave (E$_{g}$) state when electron doping from half filling to type-II vHs.

In this paper, the pairing symmetry of Ni-based transition-metal trichalcogenide superconductor is studied near type-II vHs, away from half filling 0.14eV. Within RPA pattern, we use two-sublattice two-orbital Hubbard model to calculate the pairing symmetry of 2D van der Waals (vdW) material NiPS$_{3}$. We find that the chiral even-parity $d+id$-wave (E$_{g}$) state is suppressed by odd-parity $p+ip$-wave (E$_{u}$) state when electron doping approaches the type-II vHs. In the lower doping level, chiral even-parity $d+id$-wave (E$_{g}$) is the dominant pairing in the system\cite{YXLi}. There exist pairing phase transition for this 2D superconductivity material, from singlet pairing to triplet pairing. The increasement of DOS from type-II vHs will strengthen the triplet pair. The pairing result from RPA is consistent with the analysis from spin susceptibility calculation. The peak of spin susceptibility in $\Gamma$ implies ferromagnetic fluctuation will be responsible for triplet pairing. The Fermi surface nesting from intra $\beta$ pockets promotes the instability of ferromagnetic fluctuation.

The paper is organized as follows. In Sec. II, we present the band structure, DOS and Fermi surface from two-sublattice two-orbital tight-binding model based on e$_{g}$ orbitals (d$_{xz}$ and d$_{yz}$ orbitals). We find that type II vHs are only 0.14eV above the Fermi level. In Sec. III, we show the formalism of RPA approach for superconductor pairing based on multi-orbital Coulomb interactions. In Sec. IV, we analyse the spin susceptibility and pairing symmetry when electron doping is closing to vHs. The triplet pairing $p+ip$ (E$_{u}$) is the leading superconductivity state, which is caused by ferromagnetic fluctuation. Finally, we summarize and discuss these results in Sec V.

\section{electronic structure}\label{sectionii}
The nickel phosphorous trichalcogenides compounds NiPX$_{3}$ (X=S,Se) are 2D vdW materials, which consist of layered hexagonal structures\cite{YHGu}. Each layer is constructed by $MX_{6}$ edge-shared octahedral complexes. In this octahedral environment, the five d orbitals of Ni atom are split into $t_{2g}$ and $e_{g}$ groups. The $e_{g}$ orbitals are close to half-filling while the $t_{2g}$ orbitals are fully filled. The physics near Fermi surface are mainly from the $d_{xz}$ and $d_{yz}$ orbitals. In order to capture the low-energy physics, we use two-sublattice two-orbital tight-binding model on honeycomb lattice. The corresponding tight-binding model\cite{YHGu} is given by
\begin{equation}
\emph{H}_{0}=\sum_{\bm{k}\alpha\beta}\sum_{\mu\nu\sigma}h^{\alpha\beta}_{\mu\nu}(k)c^{\dag}_{\bm{k}\alpha \mu\sigma}c_{\bm{k}\beta \nu\sigma}=\sum_{\bm{k}}\psi^\dag_{\bm{k}\sigma}h(\bm{k})\psi_{\bm{k}\sigma},
\end{equation}
with
$\psi^\dag_{\bm{k}\sigma}=(c^{\dag}_{\bm{k}A1\sigma},c^{\dag}_{\bm{k}A2\sigma},c^{\dag}_{\bm{k}B1\sigma},c^{\dag}_{\bm{k}B2\sigma})$.
Here $\alpha$, $\beta$ are the sublattice indices (A,B) and $\mu$, $\nu$ are the orbital indexes ($d_{xz}$,$d_{yz}$). $c^{+}_{\alpha\mu\sigma}$ creates a spin $\sigma$ electron with momentum $k$ in $\mu$ orbital on $\alpha$ sublattice. The matrix elements of $h^{\alpha\beta}_{\mu\nu}$(k) are provided in the Appendix of Reference [46]. It is interesting that the leading hopping parameter is third nearest neighbor (TNN) hopping term. The TNN Ni cations formed superexchange antiferromagnetic state is favored in NiPX$_{3}$ (X=S,Se) parent compounds.

In Fig. 1(a), we show the orbital resolved band dispersion ($d_{xz}$ and $d_{yz}$ orbitals) from the tight-binding model. At pristine filling, there are eight Dirac points protected by $D_{3d}$ symmetry near the Fermi level. Due to charge conversation, hole pocket and electron pocket appears at $K/2$ and $K$ respectively. Based on the two-fold rotational symmetry along $\Gamma-M$, the mixture of $d_{xz}$ and $d_{yz}$ orbitals can be found along $\Gamma-K$ and $K-M$, but not $\Gamma-M$. The strongest orbital mixture occurs near the Fermi level and also around the Dirac points ($K$ and $K/2$). In order to analyse saddle points above the Fermi surface, we calculate the corresponding DOS in Fig. 1(b). We find DOS peak above the Fermi level near 0.14 eV, which verify the existence of vHs. We plot related orbital resolved Fermi surface with $\delta=0.3$ and $\delta=0.35$ per Ni atom with respect to the half-filling in Fig. 1(c) and (d). For hexagonal symmetry in two-dimensional honeycomb lattice, there are six type-II vHs (saddle points not at TRIM points) along $\Gamma-M$ or $\Gamma-K$ direction. When changing the doping level around vHs, there accompany Lifshitz transition of Fermi surfaces. Two pockets around $K/2$ make together to become one Fermi surface in Fig. 1(c). The outer pocket around $\Gamma$ is fusing with $K$ pocket as shown in Fig. 1(d). The DOS peak is closely related with topology of Fermi surface, and also make great influence on superconductor pairing. Generally, triplet pairing could compete with singlet pairing in the system with type-II vHs\cite{HYao}, which also be verified by our RPA calculation in the following section.

\begin{figure}
\centerline{\includegraphics[width=0.5\textwidth]{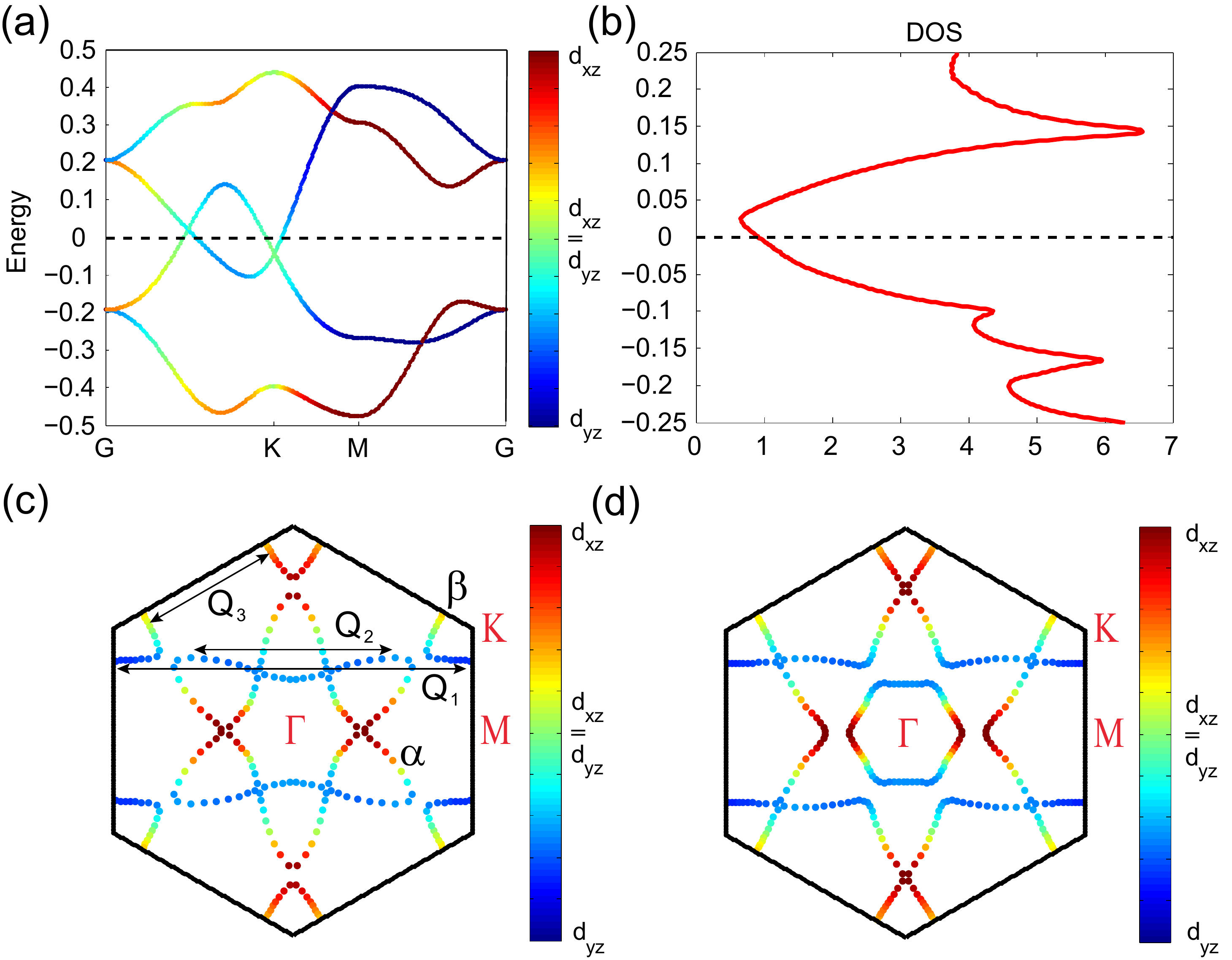}}
\caption{(color online) The orbital characters of bands are represented by different colors. (a) Band dispersion of tight-binding model based on d$_{xz}$ and d$_{yz}$ orbitals of Ni atoms. (b) The corresponding density of states (DOS) for NiPS$_3$. The
van Hove singularities (vHs) are above the Fermi level near 0.14 eV. Orbital resolved Fermi surface with two different electron fillings, $\delta=0.3$ (c) and $\delta=0.35$ (d) per Ni atom with respect to the half-filling. The type-II vHs appear along $\Gamma$-M line or $\Gamma$-K line which are not at TRIM points.}
\label{fig1}
\end{figure}

\begin{figure}
\centerline{\includegraphics[width=0.5\textwidth]{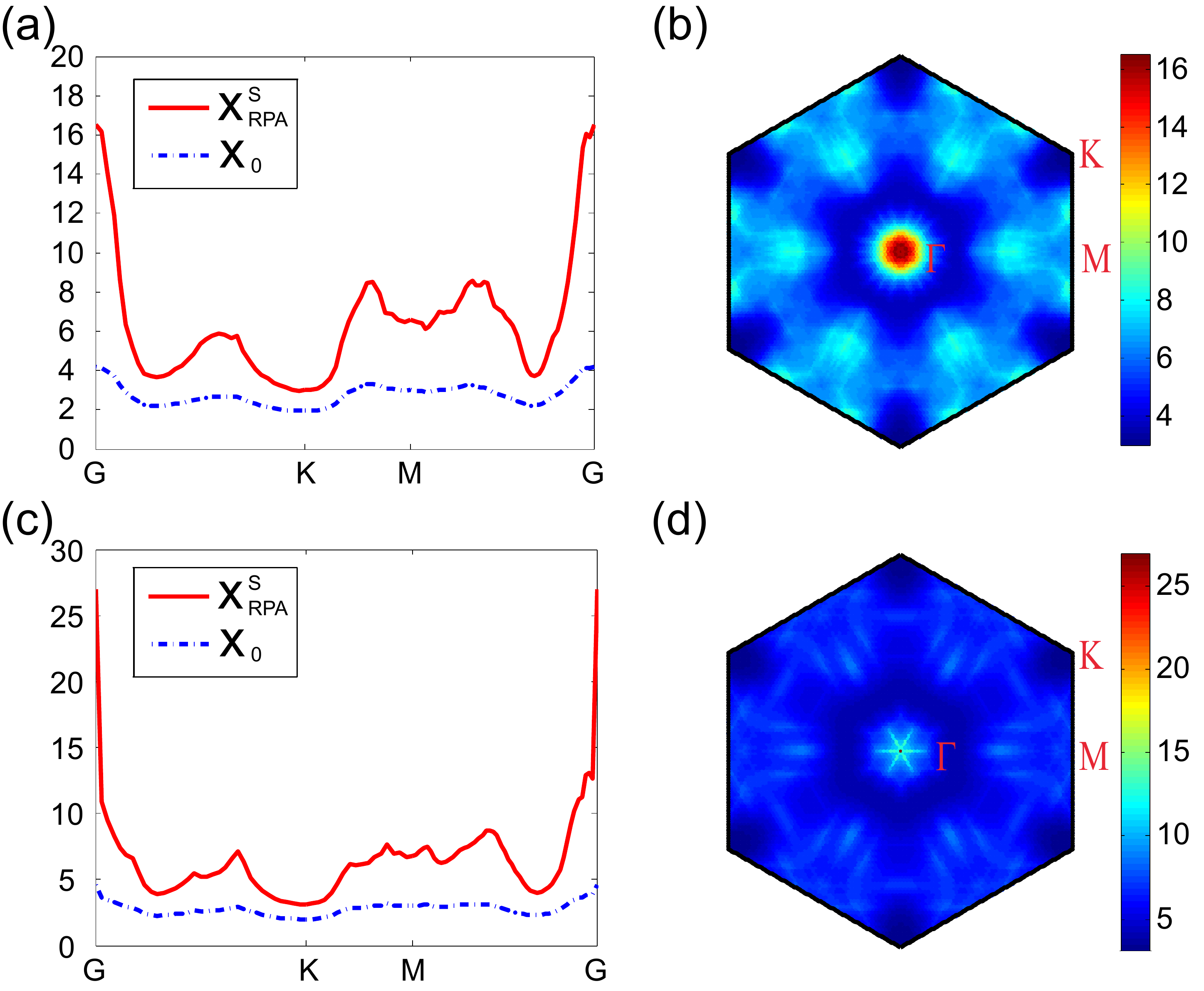}}
\caption{(color online) Susceptibilities for different electron doping. The bare susceptibilities $\chi_{0}$ and RPA spin susceptibilities $\chi^{s}_{RPA}$ along high-symmetry lines for $\delta=0.3$ (a) and $\delta=0.35$ (c) per Ni atom (near type-II vHs). The corresponding RPA spin susceptibilities in 2D Brillouin zone are shown in (b) and (d) with $U=0.3$ and $J/U=0.2$.}
\label{fig2_sus}
\end{figure}

\section{random phase approximation}\label{sectioniii}
Based on two-sublattice two-orbital tight-binding model, we consider onsite multi-orbital Hubbard interaction for superconductor pairing as
\begin{equation}
\begin{aligned}
H_{int}=&U\sum_{i,\alpha}n_{i\alpha\uparrow}n_{i\alpha\downarrow}+U'\sum_{i,\alpha<\beta}n_{i\alpha}n_{i\beta}\\
         &+J_{H}\sum_{i,\alpha<\beta,\sigma\sigma^{'}}c^{\dagger}_{i\alpha\sigma}c^{\dagger}_{i\beta\sigma^{'}}c_{i\alpha\sigma^{'}}c_{i\beta\sigma}\\
         &+J'\sum_{i,\alpha\neq\beta}c^{\dagger}_{i\alpha\uparrow}c^{\dagger}_{i\alpha\downarrow}c_{i\beta\downarrow}c_{i\beta\uparrow},
\end{aligned}
\label{hh}
\end{equation}
where $n_{i,\alpha}=n_{i,\alpha,\uparrow}+n_{i,\alpha,\downarrow}$. $U$, $U'$, $J$ and $J'$ represent the intra- and inter-orbital repulsion, the Hund's rule and pair-hopping terms. We adopt Kanamori relations $U=U'+2J$ and $J=J'$ in the next calculation, which
is required by the lattice symmetry. Considering RPA approximation\cite{Graser,Kemper}, the multi-orbital susceptibility is defined as,
\begin{eqnarray}
\chi_{l_1l_2l_3l_4}(\bm{q},\tau)&=&\frac{1}{N}\sum_{\bm{k}\bm{k}'}\langle T_{\tau} c^{\dag}_{l_3\sigma}(\bm{k}+\bm{q},\tau)c_{l_4\sigma}(\bm{k},\tau)\nonumber\\
&&c^{\dag}_{l_2\sigma'}(\bm{k}'-\bm{q},0)c_{l_1\sigma'}(\bm{k}',0) \rangle .
\end{eqnarray}
In momentum-frequency space, the multi-orbital bare susceptibility is given by
\begin{equation}
\begin{aligned}
&\chi^{0}_{l_{1}l_{2}l_{3}l_{4}}(\bm{q},i\omega_{n})=-\frac{1}{N}\sum_{\bm{k}\mu\nu}a^{l_{4}}_{\mu}(\bm{k})a^{l_{2}*}_{\mu}(\bm{k})a^{l_{1}}_{\nu}(\bm{k}+\bm{q})\times\\
&a^{l_{3}*}_{\nu}(\bm{k}+\bm{q})\frac{n_{F}(E_{\mu}(\bm{k}))-n_{F}(E_{\nu}(\bm{k}+\bm{q}))}{i\omega_{n}+E_{\mu}(\bm{k})-E_{\nu}(\bm{k}+\bm{q})},
\end{aligned}
\end{equation}
where $\mu$ and $\nu$ are the band indices, $n_{F}$ is the usual Fermi distribution, $l_{i}$ $(i=1,2,3,4)$ are the orbital indices, By diagonalizing the above multi-orbital tight-binding Hamiltonian, we obtain the $l_{i}$ orbital component of the eigenvector for band $\mu$ ($a^{l_{i}}_{\mu}(k)$) and eigenvalue $E_{\mu}(\bm{k})$. After that, we take the multi-orbital Hubbard interactions into consideration for calculating RPA susceptibilities. The corresponding RPA spin and charge susceptibilities\cite{LDZhang,XXWu5,LDZhang2} are given by
\begin{equation}
\begin{aligned}
\chi^{RPA}_{s}(\bm{q})=\chi^{0}(\bm{q})[1-\bar{U}^{s}\chi^{0}(\bm{q})]^{-1},\\
\chi^{RPA}_{c}(\bm{q})=\chi^{0}(\bm{q})[1+\bar{U}^{c}\chi^{0}(\bm{q})]^{-1},
\end{aligned}
\end{equation}
where $\bar{U}^{s}$ ($\bar{U}^{c}$) is the spin (charge) interaction matrix
\begin{displaymath}
     \bar{U}^{s/c}=
     \left(\begin{array}{cc}
           \bar{U}^{s/c}_{A}&0\\
           0&\bar{U}^{s/c}_{B}
           \end{array}
     \right),
\end{displaymath}

$$ \bar{U}^{s}_{A/B,l_{1}l_{2}l_{3}l_{4}}=\left\{
\begin{array}{rcl}
\begin{aligned}
&U                & {l_{1}=l_{2}=l_{3}=l_{4}},\\
&U'            & {l_{1}=l_{3}\neq l_{2}=l_{4}},\\
&J                & {l_{1}=l_{2}\neq l_{3}=l_{4}},\\
&J'            & {l_{1}=l_{4}\neq l_{3}=l_{2}},
\end{aligned}
\end{array} \right. $$

$$ \bar{U}^{c}_{A/B,l_{1}l_{2}l_{3}l_{4}}=\left\{
\begin{array}{rcl}
\begin{aligned}
&U                  & {l_{1}=l_{2}=l_{3}=l_{4}},\\
&-U'+2J          & {l_{1}=l_{3}\neq l_{2}=l_{4}},\\
&2U'-J           & {l_{1}=l_{2}\neq l_{3}=l_{4}},\\
&J'              & {l_{1}=l_{4}\neq l_{3}=l_{2}},
\end{aligned}
\end{array} \right. $$
In this process, we only consider the electron scattering from Fermi surfaces near the Fermi level. The effective Cooper scattering interaction is written as,
\begin{equation}
\begin{aligned}
\Gamma_{ij}(\bm{k},\bm{k}')=&\sum_{l_{1}l_{2}l_{3}l_{4}}a^{l_{2},\ast}_{\emph{v}_{i}}(\bm{k})a^{l_{3},\ast}_{\emph{v}_{i}}(-\bm{k})\\
&\times\emph{Re}\bigg[\Gamma_{l_{1}l_{2}l_{3}l_{4}}(\bm{k},\bm{k}',\omega=0)\bigg]a^{l_{1}}_{\emph{v}_{j}}(\bm{k}')a^{l_{4}}_{\emph{v}_{j}}(-\bm{k}'),
\end{aligned}
\end{equation}
where the momenta $\bm{k}$ and $\bm{k}'$ is confined in different FSs with $\bm{k}\in C_{i}$ and $\bm{k}'\in C_{j}$. The orbital vertex function $\Gamma_{l_{1}l_{2}l_{3}l_{4}}$ in spin singlet and triplet channels\cite{Graser,Kemper,XXWu5} are
\begin{equation}
\begin{aligned}
\Gamma^{S}_{l_{1}l_{2}l_{3}l_{4}}(\bm{k},\bm{k}',\omega)=&\bigg[\frac{3}{2}\bar{U}^{s}\chi^{RPA}_{s}(\bm{k}-\bm{k}',\omega)\bar{U}^{s}+\frac{1}{2}\bar{U}^{s}\\
-&\frac{1}{2}\bar{U}^{c}\chi^{RPA}_{c}(\bm{k}-\bm{k}',\omega)\bar{U}^{c}+\frac{1}{2}\bar{U}^{c}\bigg]_{l_{1}l_{2}l_{3}l_{4}},\\
\Gamma^{T}_{l_{1}l_{2}l_{3}l_{4}}(\bm{k},\bm{k}',\omega)=&\bigg[-\frac{1}{2}\bar{U}^{s}\chi^{RPA}_{s}(\bm{k}-\bm{k}',\omega)\bar{U}^{s}+\frac{1}{2}\bar{U}^{s}\\
-&\frac{1}{2}\bar{U}^{c}\chi^{RPA}_{c}(\bm{k}-\bm{k}',\omega)\bar{U}^{c}+\frac{1}{2}\bar{U}^{c}\bigg]_{l_{1}l_{2}l_{3}l_{4}},
\end{aligned}
\end{equation}
where $\chi^{RPA}_{s}$ and $\chi^{RPA}_{c}$ are the RPA spin and charge susceptibility respectively. For spin singlet and triplet
channels, the pairing vertex functions are symmetric and antisymmetric parts of the interaction, $\Gamma^{S/T}_{ij}(\bm{k},\bm{k}')=\frac{1}{2}[\Gamma_{ij}(\bm{k},\bm{k}')\pm \Gamma_{ij}(\bm{k},-\bm{k}')]$. Then, the pairing strength function for superconductor\cite{CCLiu,XXWu5,LDZhang2,LDZhang3,YTKang,FLiu} is,
\begin{equation}
\begin{aligned}
\lambda\big[\emph{g}(\bm{k})\big]=-\frac{\sum_{ij}\oint_{C_{i}}\frac{d\bm{k}_{\|}}{\emph{v}_{\emph{F}}(\bm{k})}\oint_{C_{j}}\frac{d\bm{k}'_{\|}}{\emph{v}_{\emph{F}}(\bm{k}')}\emph{g}(\bm{k})\Gamma^{S/T}_{ij}(\bm{k},\bm{k}')\emph{g}(\bm{k}')}{(2\pi)^{2}\sum_{i}\oint_{C_{i}}\frac{d\bm{k}_{\|}}{\emph{v}_{\emph{F}}(\bm{k})}\big[\emph{g}(\bm{k})\big]^{2}},
\end{aligned}
\end{equation}
where $v_{F}(\bm{k})=|\nabla_{k}E_{i}(\bm{k})|$ is the Fermi velocity on a given Fermi surface sheet $C_{i}$.

\begin{figure}
\centerline{\includegraphics[width=0.5\textwidth]{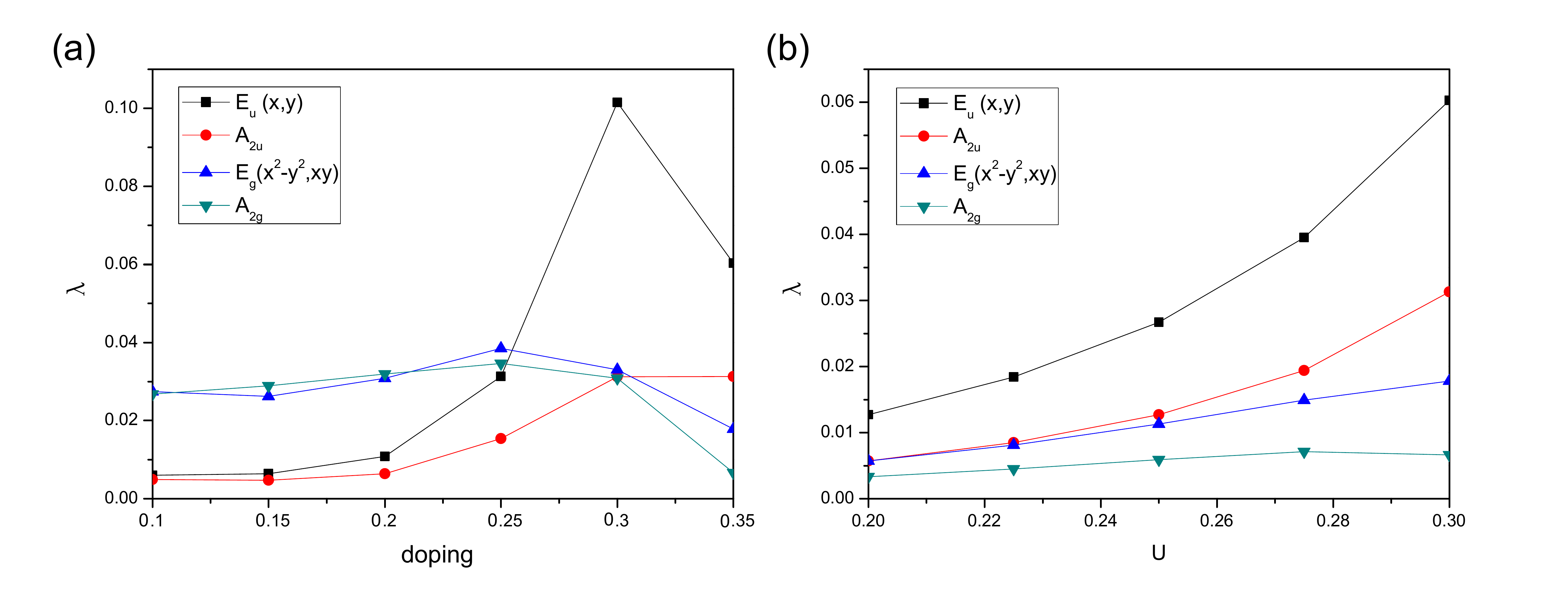}}
\caption{(color online) Leading pairing strengths in spin singlet ($E_{g}$($d$-wave),$A_{2g}$($I$-wave)) and triplet ($E_{u}$($p$-wave),$A_{2u}$($f$-wave)) channels for superconducting states with different electron dopings $\delta$ at U=0.3 and J/U=0.2 in (a). The pairing strengths with $\delta=0.35$ per Ni atom at J/U=0.2 are plotted in (b).}
\label{fig3}
\end{figure}

\begin{figure}
\centerline{\includegraphics[width=0.5\textwidth]{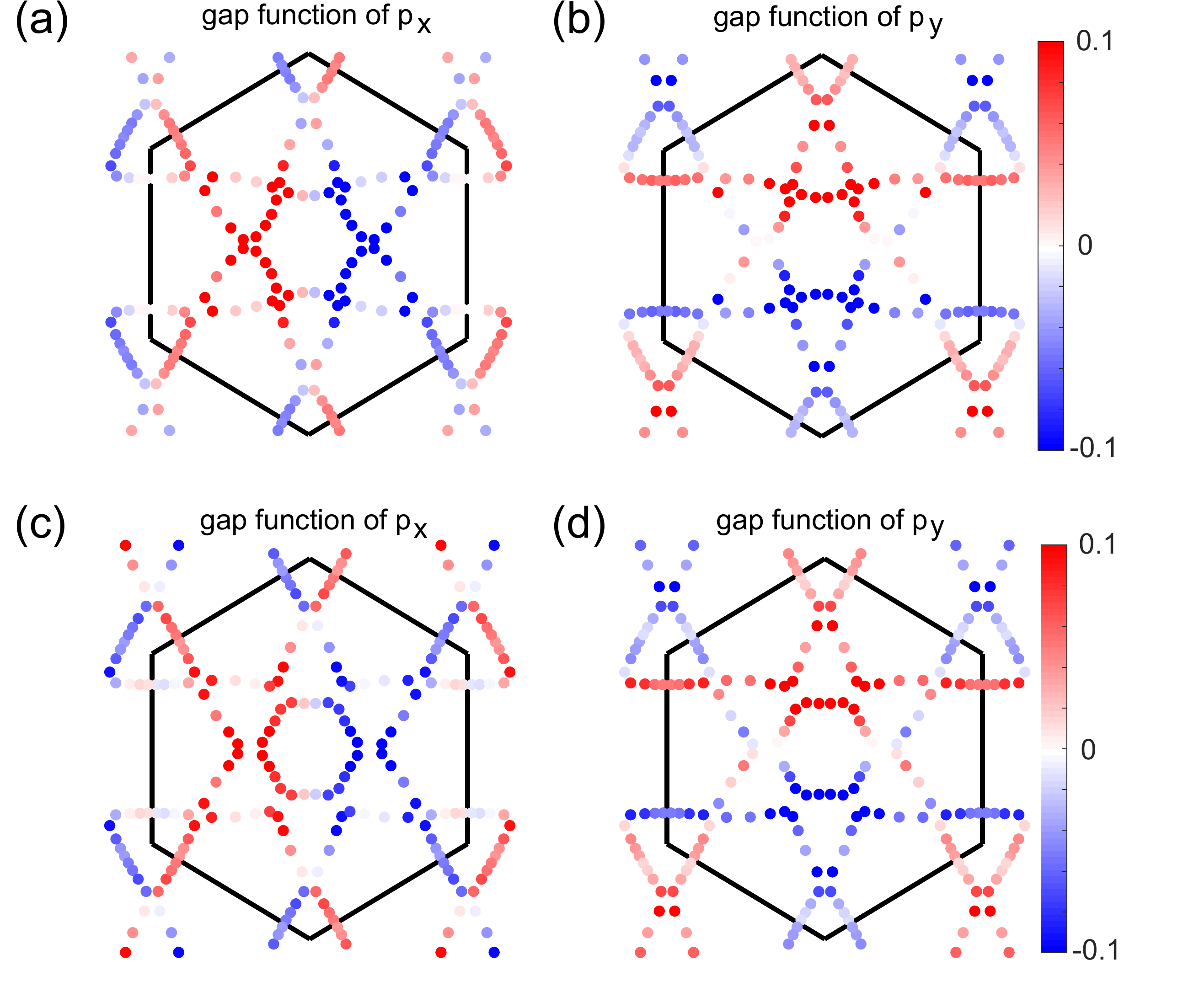}}
\caption{(color online) The gap functions $p_{x}$ and $p_{y}$ of leading triplet pairings with U=0.3 and J/U=0.2 at $\delta=0.3$ ((a) and (b)) and $\delta=0.35$ per Ni atom ((c) and (d)).}
\label{fig4}
\end{figure}

\section{susceptibility and pairing symmetry}\label{sectioniv}
Based on the multi-orbital RPA method (weak coupling approach)\cite{Graser,Kemper}, we investigate the pairing symmetry of electron doped nickel phosphorous trichalcogenides compounds NiPS$_{3}$. Due to the existence of type-II vHs peak in DOS near Fermi level, we mainly discuss the electron doping level up to $\delta=0.3$ and $\delta=0.35$ per Ni atom separately. In order to analyse the pairing symmetry results, we first calculate the bare susceptibility $\chi_{0}$ and spin susceptibility $\chi^{s}_{RPA}$ along high-symmetry lines at two different doping levels in Fig. 2(a) and 2(c). For bare susceptibility at $\delta=0.3$ per Ni atom doping level (blue dash-dot line in Fig. 2(a)), there is a prominent peak at $\Gamma$, a smooth plateau around $M$ and a broaden peak at $K/2$. The first one is mainly contributed by the intrapocket nesting between rather flat bands in NNN pockets $\beta$ (Q$_{1}$ in Fig. 1(c)). The intrapocket nesting Q$_{2}$ between NNN pockets $\alpha$ is responsible for the second peak around $M$. The third peak at $K/2$ is ascribed to the intrapocket nesting Q$_{3}$ between NN pockets $\beta$. Then, we consider the RPA spin susceptibility for superconducting instability with $U=0.3$ eV and $J/U=0.2$ in Fig. 2(a). All the mentioned above peaks are enhanced significantly. Especially, the sharp peak at $\Gamma$ near divergence indicates the ferromagnetic fluctuation between unit cells is dominant.
By checking the eigenvectors of susceptibility matrix corresponding to the largest eigenvalue, we find all the signs of eigenvectors are positive which implies ferromagnetic fluctuation in a unit cell. Compared with lower doping level at $\delta=0.1$ per Ni atom, the emergence of spin susceptibility peak at $\Gamma$ means that the ferromagnetic fluctuation is competing with antiferromagnetic fluctuation. Undoubtedly, the type-II vHs could strengthen the ferromagnetic fluctuation. In order to make the spin susceptibility clearly, we plot the corresponding susceptibility in 2D Brillouin zone. From 2D pattern in Fig. 2(b), it is clear that C$_{6}$ symmetry is maintained, a sharp peak at $\Gamma$ and another peaks around $M$. With further doping to $\delta=0.35$ per Ni atom, the peak at $\Gamma$ becomes sharper and the peaks around $M$ move toward $\Gamma$ and $K$ in Fig. 2(c) and 2(d). Due to the underestimation of interaction parameter in RPA method, we adopt intraorbital repulsive interaction parameter $U=0.3$ below the critical point $U_{c}=0.35$ (to avoid magnetic instability) and Hund's coupling $J/U\leq 0.2$.

In this section, we mainly focus on the doping level near type-II vHs. Based on the irreducible representation of $D_{3d}$ point group in this material, we classify the pairing states into subgroups of this point group. Fig. 3(a) shows the leading pairing strengths in singlet and triplet channels with different electron dopings at fixed $U=0.3$ and $J/U=0.2$. From lower doping ($\delta=0.1$) to type-II vHs ($\delta=0.35$), the system exists pairing phase transition from singlet pair to triplet pair. In the low doping level, the nearly degenerate singlet pair states $E_{g}$($x^{2}-y^{2},xy$) and $A_{2g}$ overcome the triplet pair states $E_{u}(x,y)$ and $A_{2u}$. When the doping level near type-II vHs, these singlet parings are suppressed by triplet pair state $E_{u}(x,y)$. Undoubtedly, the type-II vHs peak in DOS enhances the strength of triplet pairing. This result has also been anticipated by above RPA susceptibility analysis that the dominating ferromagnetic fluctuation favors triplet pairing. In Fig. 3(b), we plot the pairing eigenvalues as a function of U with a fixed $J/U=0.2$ with $\delta=0.35$ per Ni atom. The leading pairing state is still $E_{u}(x,y)$ and the pairing strength is increasing with increased interaction $U$. In Fig. 4, we plot the gap functions of leading two-fold degenerate pairing state $E_{u}(x,y)$ with $U=0.3$ and $J/U=0.2$ at $\delta=0.3$ ((a) and (b)) and $\delta=0.35$ ((c) and (d)) per Ni atom respectively. For $p_{x}$ ($p_{y}$) with $\delta=0.3$, the pairing nodes along y (x) axis with mirror symmetry $M_{x}$ ($M_{y}$). The gap function on $\beta$ pocket is comparable to that of $\alpha$ pocket. The superconducting orders connected by nesting vector ($Q_{1}$) between flat bands in NNN pockets $\beta$ have the same sign. For gap functions with $\delta=0.35$ in Fig. 4(c) and (d), there have the similar phenomenon in system. From above RPA calculation, type-II vHs peak in DOS induces odd-parity $p_{x}+ip_{y}$-wave ($E_{u}$) pairing state in electron doped nickel phosphorous trichalcogenides compounds NiPS$_{3}$.

\section{Conclusion}\label{sectionv}
In this paper, we have investigated the pairing symmetry for Ni-based transition metal trichalcogenide NiPS$_{3}$ based on two-sublattice two-orbital Hubbard model. By applying multi-orbital RPA method, we find that the odd-parity $p+ip$ (E$_{u}$) pairing state overcomes chiral even-parity $d+id$ (E$_{g}$) state. The enhancement of ferromagnetic fluctuation induced by type-II vHs peak in DOS is responsible for triplet pairing E$_{u}$. The nesting vector Q$_{1}$ between NNN pockets $\beta$ results in the instability peak of RPA spin susceptibility at $\Gamma$. This implies triplet pairing is the leading state which is consistent with our RPA's pairing calculation. The competition between ferromagnetic and antiferromagnetic fluctuation makes the transition from singlet to triplet pairing while doping approaches the type-II vHs. The effect of electronic DOS on unconventional superconductor's pairing may be realized in the layered Ni-based transition-metal trichalcogenides.

\end{document}